\documentclass[aps,pra,reprint,superscriptaddress,nofootinbib]{revtex4-1}
\usepackage{amsmath}
\usepackage{amssymb}
\usepackage{mathrsfs}
\usepackage{graphicx}
\usepackage[caption=false]{subfig}
\usepackage{enumitem}
\usepackage{color}
\usepackage{xcolor}
\usepackage[normalem]{ulem}
\usepackage[percent]{overpic}
\usepackage{bm}
\usepackage{rotating}
\usepackage{hyperref}
\usepackage{cancel}
\usepackage{cancel}
\usepackage{verbatim}
\usepackage{mathtools}
\usepackage{graphicx}
\usepackage{tensor}
\usepackage{verbatim}
\usepackage{MnSymbol}
\usepackage{booktabs}
\graphicspath{ {images/} }
\usepackage{braket}
\usepackage{array}

\newcolumntype{P}[1]{>{\centering\arraybackslash}p{#1}}
\newcolumntype{M}[1]{>{\centering\arraybackslash}m{#1}}
\newcommand{\ii}{\mathrm{i}}

\renewcommand{\d}{\mathrm{d}}

\newcommand{\be}{\begin{equation}}
\newcommand{\bel}[1]{\begin{equation}\label{#1}}
\newcommand{\ee}{\end{equation}}

\begin{document}
\title{A classification of Markovian fermionic Gaussian master equations}

\author{Marvellous Onuma-Kalu}
\email{monumaka@uwaterloo.ca}
\affiliation{Dept. of Physics and Astronomy, University of Waterloo, Waterloo, ON, N2L 3G1, Canada}

\author{Daniel Grimmer}
\email{dgrimmer@uwaterloo.ca}
\affiliation{Institute for Quantum Computing, University of Waterloo, Waterloo, ON, N2L 3G1, Canada}
\affiliation{Dept. of Physics and Astronomy, University of Waterloo, Waterloo, ON, N2L 3G1, Canada}

\author{Robert B. Mann}
\email{rbmann@uwaterloo.ca}
\affiliation{Dept. Physics and Astronomy, University of Waterloo, Waterloo, ON, N2L 3G1, Canada}
\affiliation{Institute for Quantum Computing, University of Waterloo, Waterloo, ON, N2L 3G1, Canada}
\affiliation{Perimeter Institute for Theoretical Physics, Waterloo, ON, N2L 2Y5, Canada}

\author{Eduardo Mart\'{i}n-Mart\'{i}nez}
\email{emartinmartinez@uwaterloo.ca}
\affiliation{Institute for Quantum Computing, University of Waterloo, Waterloo, ON, N2L 3G1, Canada}
\affiliation{Dept. Applied Math., University of Waterloo, Waterloo, ON, N2L 3G1, Canada}
\affiliation{Perimeter Institute for Theoretical Physics, Waterloo, ON, N2L 2Y5, Canada}

\begin{abstract}
We introduce a classification scheme for the generators of open fermionic Gaussian dynamics. We simultaneously partition the dynamics along the following four lines: 1) unitary vs. non-unitary, 2) active vs. passive, 3) state-dependent vs. state-independent, and 4) single-mode vs. multi-mode. We find that only nine of these sixteen types of dynamics are possible. Using this partition we discuss the consequences of imposing complete positivity on fermionic Gaussian dynamics. In particular, we show that completely positive dynamics must be either unitary (and so can be implemented without an quantized environment) or active (and so must involve particle exchange with an environment).
\end{abstract}

\maketitle

\section{Introduction}
 An open quantum system is one whose dynamics are influenced by its interaction with its surroundings. Because most realistic quantum systems are open, understanding and controlling the dynamics that arise due to system-environment interaction is critical to a wide variety of experimentally and theoretically interesting scenarios \cite{Plenio1999,Plenio2002,Chin2012,Huelga1997,Rivas2010}. For instance, analyzing how a system behaves when coupled to a heat bath is central to many models of thermodynamics as well as to everyday laboratory scenarios. There are several approaches to study the dynamics of open quantum systems, each depending on the specific quantum system in question and the demand thereof.

 One powerful tool in simplifying the description of open quantum systems is Gaussian quantum mechanics (GQM)  \cite{Weedbrook2012,D2005}, a subtheory of quantum mechanics. The theoretical concepts of GQM includes gaussian states and Gaussian transformations (those that take Gaussian states to Gaussian states). Gaussian states and transformations have simple mathematical structure and can be easily produced in the laboratory. As a result, GQM has been applied in areas including quantum information processing \cite{Weedbrook2012,Eisler2015,Eliska2018,Kraus2009}, quantum computing \cite{D2005,Bravyi2000,B2005,serge2011,PhysRevB.88.035121,PhysRevB.94.045316}, quantum entanglement \cite{Abotero2003,botero2004,isert2018,Richter2017}, thermodynamics \cite{PhysRevLett.120.190501,PhysRevA.90.020302,PhysRevA.90.062329,Melo_2013} and quantum thermodynamics \cite{eric2016,marvy2018,campbell2015}.

 GQM was recently used to study the dynamics of open bosonic Gaussian systems \cite{Dan2018}. In particular the bosonic Gaussian master equation was partitioned along the following four lines simultaneously: 1) unitary vs. non-unitary  (whether or not the interaction requires a quantized environment), 2) active vs. passive  (whether or not particles are exchanged with the environment), 3) state-dependent vs. state-independent, and 4) single-mode vs. multi-mode  (whether or not the interaction couples the various modes to each other). Of the sixteen potential types of bosonic dynamics, only eleven  were found to be possible \cite{Dan2018}. Following this partition, connections were made between the complete positivity of the dynamics and its ability to allow for the flow of quantum information and of particles between the system and its environment.
  
In this paper we  seek to construct the partition of fermionic Gaussian master equation \cite{serge2011} analogous to the bosonic case considered in \cite{Dan2018}. In particularly we will show that for fermionic systems, only nine of the potential sixteen types of dynamics are possible. We note that the dynamics of fermionic Gaussian systems appear to be more restricted than that for bosonic Gaussian systems. Additionally imposing complete positivity, we find that the presence of any non-unitary dynamics necessitates the presence of a minimum amount of noise. Since this noise is active, any non-unitarity implies particle flux with the environment.

Our paper is organized as follows. In Sec.\ref{notepre}, we introduce the basic concept of fermionic systems. In Sec. \ref{Characterizing}, We give a detailed discussion on the different dynamics that the system's master equation can produce and explicitly partition these dynamics according to their utility.  In Sec. \ref{Naming}, we present simple examples of each type of dynamics produced by the partition.  We close with conclusions summarizing our work and discuss some future directions.

\section{A Review of Fermionic Systems}\label{notepre}

In this section we review fermionic systems in general and fermionic Gaussian systems in particular. We do this for the convenience of the reader and to establish our notation. For additional resources on these topics see \cite{Bravyi2000}.

\subsection{The Algebra of Fermions}

Consider a system of $N$ fermionic modes each described by its creation and annihilation operators, $\hat{a}^{\dagger}_j$ and $\hat{a}_j$, where $j=1,2,\cdots,N$ labels the system's modes. Since the modes are fermionic, these operators obey the canonical anti-commutation relations,
\begin{align}\label{CanonAntiComm1}
\lbrace \hat{a}_{i},\hat{a}_{j}^{\dagger}\rbrace = \delta_{ij} \, \hat{\openone}, \quad \lbrace \hat{a}_{i},\hat{a}_{j}\rbrace = 0 = \lbrace \hat{a}_{i}^{\dagger},\hat{a}_{j}^{\dagger}\rbrace,
\end{align}
where $\hat{\openone}$ is the identity operator, $\delta_{ij}$ is the Kronecker delta and $\{\hat{A},\hat{B}\}\coloneqq\hat{A}\hat{B}+\hat{B}\hat{A}$ is the anti-commutator.  The free Hamiltonian for these modes is
\bel{FermionicFreeHam}
\hat{H}_0=\sum_{j=1}^N E_j \, \hat{n}_j
\ee
where $E_j$ is this $j^{th}$ mode's excitation energy and \mbox{$\hat{n}_j=\hat{a}_j^\dagger\hat{a}_j$} is the number operator for the $j^{th}$ mode.

These fermionic modes can be equivalently described in terms of their Majorana operators, defined as,
\begin{align}
\hat{x}_{j} 
\coloneqq (a_{j} + a^{\dagger}_{j})/\sqrt{2}, \quad \hat{p}_{j} 
\coloneqq \ii (a_{j} - a_{j}^{\dagger})/\sqrt{2}.
\end{align}
Note that these operators are Hermitian. Written in terms of the Majorana operators the canonical anti-commutation relations are,
\begin{align}
\lbrace \hat{x}_{i}, \hat{x}_{j}\rbrace = \lbrace \hat{p}_{i}, \hat{p}_{j}\rbrace = \delta_{ij} \, \hat{\openone}, \quad \lbrace \hat{x}_{i}, \hat{p}_{j}\rbrace = 0,
\end{align}
and the number operator for the $j^{th}$ mode is \mbox{$\hat{n}_j=\frac{1}{2} - \frac{\ii}{2}[\hat{x}_j,\hat{p}_j]$}. 

We can collect the Majorana operators into an 2N-dimensional operator-valued vector,
\begin{align}\label{vomoperators}
\hat{\mathbf{r}} = (\hat{x}_{1},\hat{p}_{1},\hat{x}_{2},\hat{p}_{2},\cdots,\hat{x}_N,\hat{p}_N)^{\intercal}.
\end{align}
Thinking of the Majorana operators as somewhat analogous to the position and momentum operators for bosonic modes, we can think of this vector as defining a phase space for the fermionic modes. In terms of this vector {\eqref{vomoperators}} the canonical anti-commutation relations further simplify to
\begin{align}\label{oform}
\lbrace \hat{r}_{n},\hat{r}_{m}\rbrace = \delta_{nm} \, \hat{\openone},
\end{align}
where $n,m=1,\dots,2N$. The number operator for the $j^{th}$ mode is \mbox{$\hat{n}_j=\frac{1}{2} - \frac{\ii}{2}[\hat{r}_{2j-1},\hat{r}_{2j}]$}.

\subsection{Physical fermionic states}
Using this operator algebra, one can define various states for the system by applying combinations of creation operators on the vacuum state, $\ket{0}$, which is annihilated by each mode's annihilation operator, $\hat{a}_j\ket{0}=0$ for all $j$. For instance in a system with $N=3$ modes we have $\ket{001}+\ket{110}\coloneqq (\hat{a}^\dagger_3+\hat{a}^\dagger_1\hat{a}^\dagger_2)\ket{0}$. However, as we will see, the scope and physicality of such constructions are limited due to the algebraic properties of fermions discussed above.

First, we note that by taking $i=j$ in \eqref{CanonAntiComm1} we have $\hat{a}^\dagger_j{}^2=0$. This implies that each mode may have at most one excitation (a manifestation of the Pauli exclusion principle). Note that this implies that $0\leq\langle\hat{n}_j\rangle\leq1$.

Second, we note that fermionic excitations behave non-trivially when rotated by $2\pi$  around any axis. Specifically they pick up a sign change,
\be
\hat{U}_{2\pi}\ket{0} = \ket{0}\!,
\ 
\hat{U}_{2\pi}\ket{01} = - \ket{01}\!,
\ 
\hat{U}_{2\pi}\ket{11} = (-1)^2 \ket{11}, 
\ee
relative to unexcited states. Note that the effect that $\hat{U}_{2\pi}$ has on elements of the Fock basis is to flip their sign if the total number of excitations is odd. That is, $\hat{U}_{2\pi}$ is just the parity operator, $\hat{P}$. Since rotation by $2\pi$ should not change the state, we can identify physical states as those whose state vector only changes by a global phase when acted on by $\hat{U}_{2\pi}=\hat{P}$,
\begin{align}
\hat{P}\ket{\psi} = e^{\ii \theta} \ket{\psi}.
\end{align}
As such, certain superpositions of fermionic excitations are unphysical \cite{Aharonov1967}. For instance the state, $\ket{0} + \ket{1}$, is unphysical because
\begin{align}
\hat{P}(\ket{0} + \ket{1}) = \ket{0} - \ket{1} \neq e^{\ii\theta} (\ket{0} + \ket{1}),
\end{align}
whereas $\ket{00} + \ket{11}$ and $\ket{01} + \ket{10}$ are physical states. 

Generally, pure states are physical if they are either superpositions of states with an odd number of excitations, or of states with an even number of excitations. A density matrix, $\rho$, is physical if and only if it commutes with $\hat{P}$. For example, taking $N=1$ we have $\hat{P}=\hat{\sigma}_z$ such that all physical states are incoherent in the basis $\{\ket{0},\ket{1}\}$. In other words for one  mode ($N=1$) the only physical states are thermal states  with respect to their free Hamiltonian \eqref{FermionicFreeHam}. 
More generally any  self-adjoint operator represents a physical observable if and only if it commutes with  $\hat{P}$. We note that $\hat{x}_j$ and $\hat{p}_j$ are unphysical, whereas $\hat{n}_j=\hat{a}^\dagger_j\hat{a}_j$ is physical. Indeed the mapping between a quantum state of fermions to what we usually refer to as `modes', although described by the well known Jordan-Wigner transformation \cite{JordanWignerRef}, is nonetheless non-trivial. It is in general not even possible  to carry out this operation in the same way as for bosonic systems; this has been a subject of recent debate \cite{Fermion1,Fermion2,Fermion3,Fermion4}.

\subsection{Fermionic Gaussian States and Transformation}
Now that we have reviewed the algebraic structure and physical states of fermionic systems, we can now review fermionic Gaussian systems. 

\subsubsection{Fermionic Gaussian States}
As their name suggests, fermionic Gaussian states are fully characterized by their first and second moments in the Majorana operators. The first moments of physical states vanish, $\langle\hat{r}_m\rangle=0$ (since \mbox{$\hat{P}\, \rho \, \hat{P}^\dagger=\rho$} but \mbox{$\hat{P}\, \hat{r}_m \,  \hat{P}^\dagger=-\hat{r}_m$}). Moreover, the second moments, $\langle \hat{r}_n\hat{r}_m\rangle$, have their symmetric part fixed by \eqref{oform}. Thus a fermionic Gaussian state is ultimately determined by antisymmetric parts of its second moments \cite{gmasterthesis}, which we can collect together in  a  covariance matrix
\begin{align}\label{cm}
\Gamma_{nm} \coloneqq \ii\langle [\hat{r}_{n}, \hat{r}_{m}]\rangle,
\end{align}
where we note that $\Gamma$ is $2N\times 2N$ dimensional, real-valued, and antisymmetric. 

A Gaussian state with covariance matrix $\Gamma$ corresponds to a physical state if and only if it obeys the positivity condition \cite{Bravyi2000,botero2004}
\begin{align}\label{conditn}
\ii  \Gamma \leq \openone_{2N}
\quad \text{or equivalently} \quad
\Gamma \Gamma^{\intercal} \leq \openone_{2N}.
\end{align}
Note that $A\leq B$ here means that $B-A$ is positive semi-definite and $\openone_{2N}$ is the $2N$ dimensional identity matrix. This condition guarantees that the density matrix corresponding to $\Gamma$ is positive semi-definite. As we will see, \eqref{conditn} can be interpreted as enforcing the Pauli exclusion principle. 
 
To help us see how to interpret the covariance matrix, let us first look at one  mode ($N=1$). In this case $\Gamma$ is $2\times 2$, real-valued, and antisymmetric, so it must be of the form
\be\label{gamform}
\Gamma
=\nu \, \omega;
\qquad
\omega
=\begin{pmatrix} 
0 & 1\\ 
-1 & 0
\end{pmatrix},
\ee
for some real parameter $\nu$. As we discussed above, for $N=1$ all physical states are thermal states  with respect  to their free Hamiltonian \eqref{FermionicFreeHam}. Thus we can interpret $\nu$ as a temperature monotone. Calculating the expected population of this mode we find
\be
\langle\hat{n}_1\rangle
=\frac{1}{2}-\frac{1}{2}\langle\ii[\hat{r}_1,\hat{r}_2]\rangle
=\frac{1}{2}-\frac{1}{2}\Gamma_{12}
=\frac{1}{2}-\frac{\nu}{2}.
\ee
In this case the positivity condition, \eqref{conditn}, requires that $\openone_2-\ii\nu\omega$ has nonnegative eigenvalues. This restricts $\nu$ to have $-1\leq\nu\leq1$ such that $0\leq\langle\hat{n}_1\rangle\leq1$. We can determine the state's inverse temperature, $\beta$, from thermal detailed balance to be 
\be
\exp(-\beta\, E_1)
=\frac{\langle\hat{n}_1\rangle}{1-\langle\hat{n}_1\rangle}
=\frac{1-\nu}{1+\nu},
\ee
or equivalently $\nu
=\text{tanh}(\beta\, E_1/2)$ where $E_1$ is the mode's excitation energy. Note that $\nu=1$ corresponds to the ground state with $\langle\hat{n}_1\rangle=0$ and $T=0$. Decreasing $\nu$ increases the temperature until at $\nu=0$ the mode is at $T=\infty$. For $\nu<0$ the population is inverted.

To see how the covariance matrix captures correlations between modes let us now look at $N=2$ modes. In this case, we can decompose $\Gamma$ into $2$ by $2$ blocks as 
\begin{align}\label{Gdecomp}
\Gamma = \begin{pmatrix} \Gamma_1 & \gamma_{12}\\ -\gamma_{12}^\intercal & \Gamma_2 \end{pmatrix}.
\end{align}
Note $\Gamma_1$ and $\Gamma_2$ describe the reduced state of each mode. As discussed above this amounts to specifying the temperature of each mode: $\Gamma_1=\nu_1\,\omega$ and $\Gamma_2=\nu_2\,\omega$. The block-off-diagonal  terms, $\gamma_{12}$, describe the correlation between the two modes.

\subsubsection{Gaussian Unitary Transformations}

In order to apply the Gaussian formalism to some dynamic scenario, one must ensure that the relevant states not only are initially Gaussian but remain Gaussian throughout their evolution. Thus the dynamics of the system must also be Gaussian, that is it must take Gaussian states to Gaussian states.

Unitary evolution under a Hamiltonian that  is quadratic in the Majorana operators is Gaussian. In fact, all Gaussian unitary transformations for fermions are generated by some quadratic Hamiltonian \cite{gmasterthesis,serge2011}. Any such Hamiltonian can be written as
\begin{align}\label{effect}
\hat{H} = \frac{\ii}{2} \bm{\hat{r}}^{\intercal}  \mathbf{H} \bm{\hat{r}},
\end{align}
for some $2N$ by $2N$ matrix $\mathbf{H}$. Note that any terms that are linear in the Majorana operator are unphysical and any constant terms may be removed by adding a constant offset to our energy scale. Moreover if $\mathbf{H}$ has any symmetric part this can be reduced to a term proportional to $\hat{\openone}$ using \eqref{oform} and subsequently removed. Thus we can without loss of generality take $\mathbf{H}$ to be antisymmetric. In order for $\hat{H}$ to be Hermitian it is then necessary that $\mathbf{H}$ is real-valued. 

For example, taking $N=1$ we find that $\mathbf{H}$ must be of the form,
\be
\mathbf{H}
=-E \,\omega;
\qquad
\omega
=\begin{pmatrix} 
0 & 1\\ 
-1 & 0
\end{pmatrix},
\ee
for some energy scale $E$. The associated Hamiltonian, 
\begin{align}
\hat{H}
&=\frac{-\ii}{2}E(\hat{x}_1\hat{p}_1-\hat{p}_1\hat{x}_1)
=E(\hat{n}_1-\frac{1}{2}),
\end{align}
is (up to a constant offset) proportional to the mode's number operator. Thus we can interpret $E$ as the energy of the mode's excitation.

To see how quadratic Hamiltonians can couple multiple modes together, let us take $N=2$ and decompose $\mathbf{H}$ into 2 by 2 matrices as
\bel{Hdecomp}
\mathbf{H} = 
\begin{pmatrix} 
H_1 & H_{12}\\ 
-H_{12}^\intercal & H_2
\end{pmatrix}.
\ee
As above, the block-diagonal  terms set the local energy scales for each mode's excitation: $H_1=-E_1 \, \omega$ and $H_2=-E_2 \, \omega$. The block-off-diagonal  terms then describe the couplings between the modes.

Next let us look at how the vector of Majorana operators, $\hat{\mathbf{r}}$, evolve under such a quadratic Hamiltonian. In the Heisenberg picture this evolution is described by,
\begin{align}\label{HeisenbergEq}
\frac{\d}{\d t}\hat{\mathbf{r}} 
=\ii [\hat{H},\hat{\mathbf{r}}].
\end{align}
Note that $\hat{H}$ operates on $\hat{\mathbf{r}}$ component-wise as an operator in Hilbert space. The standard solution to this equation is
\be
\hat{\mathbf{r}}(t)=e^{\ii\, \hat{H} \, t}\hat{\mathbf{r}}(0)e^{-\ii \, \hat{H} \, t}.
\ee
Alternatively, one can attempt to describe the system's evolution as a transformation in phase space. As we show in Appendix \ref{HilbertPhaseCrossover}, using the identity
\be
[\hat{A}\hat{B},\hat{C}]
=\hat{A}\lbrace \hat{B},\hat{C}\rbrace - \lbrace \hat{A},\hat{C}\rbrace \hat{B},
\ee
and the antisymmetry of $\mathbf{H}$ one finds from \eqref{HeisenbergEq},
\begin{align}\label{rmaster}
\frac{\d}{\d t}\hat{\mathbf{r}} 
=\mathbf{H} \, \hat{\mathbf{r}}.
\end{align}
Note that $\mathbf{H}$ here acts on $\hat{\mathbf{r}}$ as a vector in phase space, producing a linear combination of its operator-valued entries. Solving the above equation yields
\be
\hat{\mathbf{r}}(t)=e^{\mathbf{H} \, t}\hat{\mathbf{r}}(0).
\ee
Since $\mathbf{H}$ is real-valued and antisymmetric, exponentiating it generates an element of the special orthogonal group, $e^{\mathbf{H} \, t}\in\text{SO}(2N)$. 

Thus we see that unitary evolution under a quadratic Hamiltonian corresponds to a (special-)orthogonal rotation in phase space,
\begin{align}\label{HPScorrespondence}
\mathbf{\hat{r}}(t)=\hat{U}_{G}\, \mathbf{\hat{r}}(0)\, \hat{U}^{\dagger}_{G} = O \,  \mathbf{\hat{r}}(0), 
\end{align}
where $U_{G} = e^{\ii \hat{H} t}$ and $O = e^{\mathbf{H}t}$ are (Gaussian) unitary and (special) orthogonal linear maps on the system's Hilbert space and phase space respectively. Since we are working primarily in the system's phase space, for the rest of the text we will call such dynamics orthogonal. Since these transformations have a unitary description in the system's Hilbert space all of the intuitions relating to unitary evolution carry over.

In order to track the evolution of a Gaussian state, we need only track the evolution of its covariance matrix. From \eqref{rmaster} we find,
\begin{align}\label{cov}
\frac{\d \Gamma}{\d t} 
&= \mathbf{H}\,\Gamma+\Gamma\,\mathbf{H}^\intercal,
\end{align}
which can be solved as
\begin{align}
\Gamma(t) = e^{\mathbf{H}t} \Gamma(0)e^{\mathbf{H}^{\intercal}t}.
\end{align}
Thus the system's covariance matrix is also updated by a (special-)orthogonal transformation,
\begin{align}\label{GammaO}
\Gamma &\longrightarrow O \, \Gamma \, O^\intercal.
\end{align}

\subsubsection{Open Gaussian Transformations}\label{OpenGQM}
The Gaussian unitary transformations discussed above are not the only transformations  preserving the Gaussianity of fermionic states. Analogously to Stinespring's dilation theorem, we can find fermionic Gaussian channels by examining the effect that Gaussian unitary transformations on a larger system have on a subspace corresponding to a reduced subsystem.

Consider a bipartite fermionic system, AB, where system A is composed of $N$ fermionic modes and system B is composed of $M$ fermionic modes. Suppose the joint system is in a Gaussian state. The bipartite covariance matrix can be divided into blocks as 
\begin{align}
\Gamma = \begin{pmatrix} \Gamma_{A} & \gamma_{AB} \\ -\gamma_{AB}^{\intercal} & \Gamma_{B} \end{pmatrix},
\end{align}
where $\Gamma_{A}$ and $\Gamma_{B}$ are $2N$ by $2N$ and $2M$ by $2M$ matrices respectively representing the reduced state of the individual systems and where $\gamma_{AB}$ is an $2N$ by $2M$ matrix recording the correlations between the two systems.  

Suppose that the two systems are initially uncorrelated, $\gamma_{AB}(0)=0$, and jointly undergo a Gaussian unitary transformation. As discussed above this can be represented by an orthogonal transformation,
\begin{align}
 O = \begin{pmatrix} O_{A} & O_{AB} \\ O_{BA} & O_{B}\end{pmatrix},
\end{align}
on the systems' joint phase space. Note that $O_{A}$,  $O_{B}$, $O_{AB}$ are not necessarily orthogonal themselves.

Applying this transformation to the joint state as in \eqref{GammaO} we can see that the reduced state of system $A$ is updated as {\cite{campbell2015}}
\begin{align*}
\Gamma_\text{A}(0)\to\Gamma_\text{A} = O_\text{A}\Gamma_\text{A}(0) O_\text{A}^{\intercal} + O_\text{AB}\Gamma_{B}(0) O_\text{AB}^{\intercal}
\end{align*}
where we note that the second term in this expression is antisymmetric. Thus in general we see that a fermionic Gaussian channel  $\Phi_G$ is defined by a $2N$ by $2N$ matrix $O_A$ and an antisymmetric $2N$ by $2N$ matrix $R$ as
\begin{align}\label{one}
 \Phi_G : \Gamma \rightarrow O_A \Gamma O_A^{\intercal} + R.
\end{align}
In order for such a Gaussian channel to be physical it must map physical states to physical states. That is if $\Gamma$ satisfies \eqref{conditn} then $ \Phi_G[\Gamma]$ should too. This is the case if and only if the channel obeys the complete positivity condition \cite{Bravyi2005}
\begin{align}\label{cptpcondition}
\ii R \leq \openone_{2N} - O_A \, O_A^{\intercal}.
\end{align}

We can find the general form of a fermionic Gaussian master equation by taking the above Gaussian channel to be differential as,
\begin{align}\label{diff_form}
 O_\text{A} = \openone_{2N} + A \, \d t, \quad R = C \, \d t,
\end{align}
for some $2N$ by $2N$ real-valued matrices $A$ and $C$, with $C$ antisymmetric. 

Substituting Eqn. \eqref{diff_form} in \eqref{one}, we can compute $\frac{\d}{\d t}\Gamma(t)$
\begin{align}\label{ode}
 {\frac{\d}{\d t}\Gamma(t) = A \Gamma(t) + \Gamma(t) A^{\intercal} + C.}
\end{align}
Note that Eqn. \eqref{ode} is an affine transformation comprising of a linear term $A \Gamma(t) + \Gamma(t) A^{\intercal}$ and an affine term $C$. Moreover note that $A$ is no longer required to be antisymmetric. In these two ways this evolution generalizes the unitary/orthogonal master equation, \eqref{cov}. In terms of the generators $A$ and $C$, the complete positivity condition \eqref{cptpcondition} is
\begin{align}\label{diffcptpcond}
A + A^\intercal+ \ii C \leq 0.
\end{align}
Any dynamics for the covariance matrix of the form \eqref{ode} satisfying \eqref{diffcptpcond} can be equivalently written as a differential equation for the states density matrix, $\rho$ in Lindblad form, 
\begin{align}\label{lindmeqtn}
\frac{\d}{\d t} \rho = -\ii [\hat{H},\rho] + \sum_{\alpha}\Big(2\hat{L}_{\alpha}\rho \hat{L}_{\alpha}^{\dagger} - \lbrace \hat{L}_{\alpha}^{\dagger} \hat{L}_{\alpha},\rho \rbrace\Big),
\end{align}
with a Hamiltonian, $\hat{H}$, and Lindblad operators, $\hat{L}_\alpha$, which are quadratic and linear in the system's Majorana operators respectively \cite{PhysRevA.87.012108,Prosen_2008,serge2011}.

\section{Characterizing Gaussian Master Equations}\label{Characterizing}
Now that we have concluded our review of fermionic Gaussian dynamics and states, we will now classify the different kinds of dynamics that the general fermionic Gaussian master equations, \eqref{ode} can produce.

In particular we will classify the different parts of open Gaussian evolution according to the following four dichotomies:
\begin{itemize}
    \item Orthogonal vs. Non-orthogonal
    \item Passive vs. Active
    \item Single-Mode vs. Multi-Mode
    \item State-Dependent vs. State-Independent
\end{itemize}
leading to sixteen potentially possible types of dynamics. The relevance of these divisions is fleshed out in the next subsections.

\subsection{Classification of Gaussian Evolution}\label{Criteria} 
\subsubsection{Orthogonal vs. Non-Orthogonal}
As we discussed above in equation \eqref{HPScorrespondence}, Gaussian unitary evolution in Hilbert space corresponds to an orthogonal  transformation in phase space. We can interpret non-orthogonal dynamics in the same way we interpret non-unitary dynamics. Particularly, we can view non-orthogonal dynamics as those dynamics which require an external system for  the system to exchange information with (i.e., to become correlated with).

This correspondence can be carried over to the system's master equation as well by noting that the general master equation \eqref{ode} reduces to  the orthogonal one (unitary evolution), \eqref{cov}, when
\begin{align}\label{eq}
C=0,
\quad\quad
A^\intercal=-A.
\end{align}
that is, when A is antisymmetric.  Thus the presence of an affine term, $C\neq0$, or a symmetric part of $A$ indicates non-orthogonal (non-unitary) evolution. Thus we can identify all of $C$ and the symmetric part of $A$ as the non-orthogonal parts of the dynamics
\begin{align}
C_\textsc{n}&\coloneqq C,\\
A_\textsc{n}&\coloneqq\frac{1}{2}\big(A+A^\intercal\big).
\end{align}
Likewise we can identify the antisymmetric part of $A$ as the orthogonal part of the dynamics,
\begin{align}
C_\textsc{o}&\coloneqq0,\\
A_\textsc{o}&\coloneqq\frac{1}{2}\big(A-A^\intercal\big).
\end{align}

As we have discussed above, dynamics of the form \eqref{ode} can be written in Lindblad form \eqref{lindmeqtn} with a quadratic Hamiltonian and linear  Lindblad operators in the system's Majorana operators. We can associate the orthogonal part of our dynamics with the unitary term $ -\ii [H,\rho] $, that is
\begin{align}\label{Heff}
\mathbf{H}_\text{eff}=A_\textsc{o},
\qquad
\hat{H}_{\text{eff}}= \frac{\ii}{2}\bm{\hat{r}}^{\intercal} A_\textsc{o} \bm{\hat{r}}
\end{align}
where we note that such both unitary and dissipative evolution of fermionic Gaussian states can be efficiently simulated \cite{serge2011}.

Similarly, the non-orthogonal part of the dynamics can be associated with Lindblad operators $\hat{L}_\alpha$ that are linear in the system's Majorana operators as follows,
\bel{LindbladOp}
-2A_\textsc{n} - \ii C_\textsc{n} =\sum_\alpha \gamma_\alpha \ \bm{\ell}_\alpha \, \bm{\ell}_\alpha^\dagger, \qquad \hat{L}_\alpha=\sqrt{\gamma_\alpha} \ \bm{\ell}_\alpha^\dagger \, \hat{\bm{r}} 
\ee
where $\gamma_\alpha$ are the eigenvalues of $-2A_\textsc{n} - \ii C_\textsc{n}$ (note that equation \eqref{diffcptpcond} guarantees these are positive) and $\bm{\ell}_\alpha \in \mathbb{C}^{2N}$ are $2N$ dimensional complex valued mutually orthogonal unit vectors. These expressions can be confirmed by comparison with Sec. II and III of \cite{PhysRevA.87.012108}.

\subsubsection{Passive vs. Active}\label{nrg}
In addition to classifying whether the dynamics are orthogonal or not, we can also characterize the dynamics by their effect on the average total excitation number (sum of the average excitation number in all the modes). Note this corresponds one to one with the effect on the system's energy if all the modes are uncoupled and have the same excitation energy.

We define dynamics as either active or passive depending on whether it changes or maintains the expected excitation number respectively. Dynamics that commute with the excitation number are called number conserving. We can interpret active dynamics as those dynamics that require an external system for the system to exchange particles with. Note that we could equally well refer to passive dynamics as number conserving and active dynamics as number non-conserving.

The expected excitation number can be written in terms of the system's covariance matrix $\Gamma$ as \cite{marvy2018},
\begin{align}
\langle \hat{n}\rangle 
&=\sum_{j=1}^n \langle \hat{n}_j\rangle
=\frac{N}{2}+\frac{1}{4}\text{Tr}(\Omega \,\Gamma).
\end{align}
where $\Omega$ is the symplectic matrix,
\begin{align}\label{SymplecticDef}
\Omega = \bigoplus_{j=1}^N \omega
=\openone_N\otimes\omega;
\quad 
\omega
=\begin{pmatrix}
0 & 1 \\
-1 & 0
\end{pmatrix} .
\end{align}
The rate of change of the expected excitation number can be computed using \eqref{ode} as,
\begin{align}\label{Activity}
\frac{\d}{\d t}\langle \hat{n}\rangle
=&\frac{1}{4}\text{Tr}\Big(\Omega \big(A \Gamma + \Gamma A^{\intercal} + C\big)\Big)\\\nonumber
=&\frac{1}{4}\text{Tr}\Big(\Omega A \Gamma + A^{\intercal}\Omega \Gamma + \Omega C \Big)\\
\nonumber
=&\frac{1}{4}\text{Tr}\Big( \big(\Omega A - (\Omega A)^{\intercal}\big)\Gamma + \Omega C\Big),
\end{align}
where we have used the cyclic property of trace and that $\Omega$ is antisymmetric. By our above definition, the dynamics is passive if \mbox{$ \frac{\d}{\d t}\langle \hat{n}\rangle = 0$} for all $\Gamma$. This is only the case if $\Omega A$ is symmetric and $\Omega C$ is traceless.

Thus we can identify the part of $A$ which is passive, $A_\textsc{p}$ (active $ A_\textsc{a}$), as that part which becomes symmetric (antisymmetric) when multiplied by $\Omega$. Specifically 
\begin{align}
\Omega A_\textsc{p} = \frac{1}{2} (\Omega A + (\Omega A)^{\intercal}),
\quad
\Omega A_\textsc{a} = \frac{1}{2} (\Omega A - (\Omega A)^{\intercal}).
\end{align}
so that $A_\textsc{p}+A_\textsc{a}=A$.
Using the symplectic identities $\Omega^{-1} = \Omega^{T} = -\Omega$ we have 
\begin{align}
A_\textsc{p} = \frac{1}{2}\big(A+\Omega A^\intercal\Omega\big),
\quad
A_\textsc{a}
=\frac{1}{2}\big(A-\Omega A^\intercal\Omega\big).
\label{AaDef}
\end{align}
 
In order to identify the active and passive parts of $C$ we must split it into parts that do and do not contribute to the trace of $\Omega C$. This is not trivial and will be discussed in greater detail once we introduce  the other dichotomies   in Sec. \ref{partition}.

\subsubsection{State-Dependent vs State-Independent}

Looking at the general fermionic Gaussian master equation, \eqref{ode}, we see that the effect that the linear terms (those involving $A$) have on the covariance matrix depends on the current state of the system, i.e. these terms are state-dependent. On the other hand the effect of the affine term (i.e. $C$) does not depend on the state of the system; it is state-independent. Thus we call $A$ the state-dependent part of the dynamics and $C$ the state-independent part.

The interpretation of state-independent dynamics is a bit nuanced and warrants further discussion. Consider a system evolving under the Lindblad master equation, \eqref{lindmeqtn}. Clearly all terms appearing in this equation must be linear in $\rho$ and so depends on $\rho$ (through multiplication). Thus it would appear that (at least written this way) all dynamics are state dependent. How can this be true if, as we discussed in Sec. \ref{OpenGQM}, a Gaussian state's covariance matrix undergoes a linear affine update. In the covariance matrix description it appears we can identify which dynamics are affine (i.e., state-independent) unambiguously, whereas there seems to be no such terms when we analyze the dyanamics in terms of the system's density matrix.

To resolve the tension in the above paragraph, we find helpful to think of an analogous situation in the Bloch sphere. Let $\bm{a}_{4}=(x_0,x_1,x_2,x_3)^\intercal$ be the projection of a two-level system's density matrix onto the Pauli basis, \{$\hat{\openone}_2/2$, $\hat{\sigma}_x/2$, $\hat{\sigma}_y/2$, $\hat{\sigma}_z/2$\}, that is
\begin{align}
\rho = \frac{1}{2}(x_{0} \, \hat{\openone}_2 + x_{1} \, \hat{\sigma}_x + x_{2} \, \hat{\sigma}_{y} + x_{3} \, \hat{\sigma}_{z})
\end{align}
where $x_{0},\,x_{1},\,x_{2},\,x_{3} \in \mathbb{R}$. Note that from the state's normalization condition, $\text{Tr}(\rho) = 1$, we know that $x_{0} = 1$ is fixed. Thus we can partition this trivial component of $\bm{a}_4$ from the others as $\bm{a}_4 = (1,\bm{a}_3)^\intercal$ where $\bm{a}_3 = (x_{1},x_{2},x_{3})^\intercal$ is the system's usual Bloch vector satisfying
\begin{align}
\rho = \frac{1}{2}(\hat{\openone}_2 + \bm{a}_3 \cdot \hat{\bm{\sigma}}).
\end{align}
Note that $\bm{a}_4\in\mathbb{R}^4$ lives in an affine space, that is, on a hyperplane displaced from the origin, whereas $\bm{a}_3\in\mathbb{R}^3$ does not.

Turning to the system's dynamics, we note that every completely positive trace preserving map $\Phi$ acting on $\rho$ (i.e., $\rho'=\Phi[\rho]$, where the prime denotes the system state at a later time) can be represented by unique $4 \times 4$ real-valued matrix, $M$, acting on $\bm{a}_4$ as $\bm{a}_4'=M\bm{a}_4$. Again separating out the trivial component of $\bm{a}_4=(1,\bm{a}_3)^\intercal$ we can write $M$ as,
\begin{align}
    M = \begin{pmatrix}
    1&\bm{0}\\\bm{t}& \text{T}
    \end{pmatrix},
\end{align} 
where $T$ is a $3 \times 3$ real-valued matrix and $\bm{t}$ is a real-valued 3 dimensional vector. Note that the first row of $M$ is fix by our assumption that $\Phi$ is trace preserving. We can write this dynamics in terms of the system's Bloch vector, $\bm{a}_3$, as
\begin{align}
\begin{pmatrix}
1 \\
\bm{a}_3'
\end{pmatrix}
=\bm{a}_4'
=M\bm{a}_4
=\begin{pmatrix}
1&\bm{0}
\\\bm{t}& \text{T}
\end{pmatrix}
\begin{pmatrix}
1 \\
\bm{a}_3
\end{pmatrix}
=\begin{pmatrix}
1 \\
T\bm{a}_3+t
\end{pmatrix}.
\end{align}
such that $\bm{a}_3$ obeys an linear-affine update equation \mbox{$\bm{a}_3'=T\bm{a}_3+t$}. Contrast this with the linear update equation, $\bm{a}_4' = M\bm{a}_4$ which $\bm{a}_4$ obeys. In effect we have exchanged the ``affine-ness'' of the vector's space for ``affine-ness'' in its update equation.

Analogously to the fermionic Gaussian case, we can identify the $3\times 3$ matrix $T$ as the state dependent part of the dynamics and the 3 dimensional vector $t$ as the state independent part. In this context their interpretations is clear. Viewing $T$ and $\bm{t}$ as subblocks of the $4\times 4$ update map $M$ we can see that $T$ describes how the $\sigma_{x,y,z}$ components of $\bm{a}_4$ rotate into the $\sigma_{x,y,z}$ components of $\bm{a}_4'$. Likewise we can see that $\bm{t}$ describes how the trivial identity component of $\bm{a}_4$, i.e. $x_0$, rotate into the $\sigma_{x,y,z}$ components of $\bm{a}_4'$. Since all valid states have the same value for the identity component of $\bm{a}_4$ the dynamics coming from $\bm{t}$ is in a sense state independent.

Moreover we should note that $\bm{t}$ also describes the non-unital part of the dynamics, that is the dynamics which displaces the maximally mixed state, $\rho=\hat{\openone}_2/2$ or equivalently $\bm{a}_3=0$. Concretely, $\bm{t}=\frac{\d}{\d t}\bm{a}_3$ at $\bm{a}_3=0$. For finite dimensional systems the presence of non-unital dynamics \cite{Lidar2006} is necessary and sufficient for the dynamics to be able to increase some state's purity.

In the fermionic Gaussian case, the role of the trivial component of $\bm{a}_4$ is played by the symmetric part of the state's second moments, $\langle \hat{r}_n\hat{r}_m\rangle$, which are fixed by the canonical anti-commutation relations, \eqref{oform}. We should also note that as in our Bloch sphere example, the fermionic state-independent dynamics, corresponds to the non-unital part of the dynamics. This can be seen by noting that the fermionic maximally mixed state is here represented by the infinite temperature state, $\Gamma=0$. Analogously to our Bloch sphere example, we can then identify the state independent part of the dynamics as $C=\frac{\d}{\d t}\Gamma$ at $\Gamma=0$. As we will see $C$ is indeed associated with purifying dynamics.

\subsubsection{Single-Mode vs. Multi-Mode}\label{SingleModeMultiMode}
Recall that in our definition of $\bm{\hat{r}}$, \eqref{vomoperators}, the Majorana operators corresponding to each mode are listed adjacent to each other, in pairs. From this it follows that in the various matrices defined from and acting on $\bm{\hat{r}}$ (i.e., $\Gamma$, $\mathbf{H}$, $A$, $C$) adjacent pairs of rows and columns correspond to individual modes. Dividing these matrices into 2 by 2 blocks one finds that the block diagonal terms describe the correlations within each of the modes and how they are coupled to themselves, whereas the block-off-diagonal  terms describe the multi-mode correlations and how they couple to each other.

Thus we can think of the block-diagonal parts of both $A$ and $C$ as being responsible for the single-mode dynamics of the system while the block-off-diagonal parts of $A$ and $C$ describe the couplings between the modes, i.e. the multi-mode dynamics. We can interpret multi-mode dynamics as those dynamics that couple different modes of the system together either directly or indirectly through an environment/third party.

Note that if one chooses to define  fermionic modes in a different way then different dynamics will be considered single-mode and multi-mode. Similarly, these new modes would have their own number operators and therefore which dynamics are considered active/passive will also change. Note however that which dynamics are (non-)orthogonal and which are state (in-)dependent is independent of how we define our modes.

Moreover, all aspects of our classification scheme are invariant under a local change of basis within each mode.

Finally we note that instead of partitioning the system all the way down to individual modes, one could instead consider a more general partition into subsets of modes. This straightforwardly generalizes the notion of multi-mode dynamics to mutli-partite dynamics. These collections of modes could, for instance, correspond to those accessible by spatially separated parties.

\subsection{Partitioning the Gaussian master equations}
\label{partition}

Now that we have discussed four ways of partitioning fermionic Gaussian dynamics we will explicitly perform all four partitions at once. We have summarized the results of this partition in Table \ref{SUAPTable}.

As discussed above, it is trivial to partition the dynamics into its state-dependent and state-independent parts (A and C respectively). We will now divide each of these into its orthogonal active (OA), orthogonal passive (OP), non-orthogonal active (NA) and non-orthogonal passive (NP) parts. Then we will separate the single-mode (S) dynamics from the multi-mode dynamics (M).

As discussed above, finding the orthogonal and non-orthogonal parts of $A$ involve finding its symmetric and antisymmetric parts. Similarly finding the active and passive parts of $A$ involve symmetrizing and antisymmetrizing $\Omega A$. To find the orthogonal active part of A, for instance, we do both antisymmetrizations consecutively, 
\begin{align}
A_\textsc{oa}
&\coloneqq\frac{1}{2}(A_\textsc{a}-A_\textsc{a}^\intercal)
=\frac{1}{2}\Omega^{-1}(\Omega A_\textsc{o}-(\Omega A_\textsc{o})^\intercal).
\end{align}
Note that performing the antisymmetrizations in either order and expanding yields,
\begin{align}
A_\textsc{oa}
&=\frac{1}{4}(A-A^\intercal-\Omega A^\intercal\Omega+\Omega A\Omega).
\end{align}
One can quickly confirm that both $A_\textsc{oa}$ and $\Omega A_\textsc{oa}$ are antisymmetric as desired.

Similarly we can identify the other parts of $A$:
\begin{align}
A_\textsc{op}
&\coloneqq\frac{1}{2}(A_\textsc{p}-A_\textsc{p}^\intercal)
=\frac{1}{2}\Omega^{-1}(\Omega A_\textsc{o}+(\Omega A_\textsc{o})^\intercal),\\
A_\textsc{na}
&\coloneqq\frac{1}{2}(A_\textsc{a}+A_\textsc{a}^\intercal)
=\frac{1}{2}\Omega^{-1}(\Omega A_\textsc{n}-(\Omega A_\textsc{n})^\intercal),\\
A_\textsc{np}
&\coloneqq\frac{1}{2}(A_\textsc{p}+A_\textsc{p}^\intercal)
=\frac{1}{2}\Omega^{-1}(\Omega A_\textsc{n}+(\Omega A_\textsc{n})^\intercal).
\end{align}

To find more convenient expressions for these, it is helpful for us to introduce a basis for the real $2$ by $2$ matrices
\bel{2by2basis}
\openone_2
=\begin{pmatrix}
1 & 0 \\
0 & 1 \\
\end{pmatrix},
\,
\omega=\begin{pmatrix}
0 & 1 \\
-1 & 0 \\
\end{pmatrix},
\, 
X=\begin{pmatrix}
0 & 1 \\
1 & 0 \\
\end{pmatrix},
\,
Z=\begin{pmatrix}
1 & 0 \\
0 & -1 \\
\end{pmatrix},
\ee
and to expand $A$ over this basis in the second tensor factor   
\begin{align}\label{expansiona}
A\,= \, &\,A_1 \otimes\openone_2 + A_w \otimes\omega + A_x \otimes X + A_z \otimes Z 
\end{align}
where $A_\mu$ are $N$ by $N$ matrices for $\mu\in\{1,w,x,z\}$. 

As an example, let us again calculate the orthogonal active part of $A$ now using this expansion. Defining the linear functions,
\begin{align*}
\text{anti}(A) \coloneqq \frac{1}{2}(A - A^{\intercal}), \quad \text{and} \quad \text{sym}(A) \coloneqq \frac{1}{2}(A + A^{\intercal}),
\end{align*}
we can compute the orthogonal part of $A$ as, 
\begin{align}\nonumber
A_\textsc{o} =& \text{anti}(A)\\\nonumber
= & \text{anti} (A_1 \otimes\openone_2 + A_w \otimes\omega + A_x \otimes X + A_z \otimes Z)\\\nonumber
=& \text{anti} (A_1 \otimes\openone_2 ) + \text{anti}(A_w \otimes\omega )\\\nonumber
&+ \text{anti}(A_x \otimes X )+ \text{anti}( A_z \otimes Z).
\end{align}
Using the identities,
\begin{align}
\text{anti}( A_\mu \otimes Y)
&=\text{anti}(A_\mu) \otimes Y 
\quad\text{if $Y$ is symmetric,}\\
\nonumber
\text{anti}( A_\mu \otimes Y)
&=\text{sym}(A_\mu) \otimes Y
\quad\text{if $Y$ is antisymmetric,}
\end{align}
this simplifies to,
\begin{align}
A_\textsc{o} =& \text{anti}(A_{1})\otimes \openone_{2} + \text{sym}(A_{\omega}) \otimes \omega \\
\nonumber
&+ \text{anti}(A_{x})\otimes  X + \text{anti}(A_{z})\otimes Z.    
\end{align}
The orthogonal and active part of $A$,
\begin{align}
A_\textsc{oa} =& \Omega^{-1}\text{anti}(\Omega A_\textsc{o}),  
\end{align}
can be calculated by recalling from \eqref{SymplecticDef} that \mbox{$\Omega=\openone_N\otimes\omega$} and noting that multiplying on the left by $\omega$ cycles the 2 by 2 basis as,
\bel{2by2cycle}
\omega:
\quad\openone_2\to\omega,
\quad\omega\to-\openone_2,
\quad X\to Z,
\quad Z\to-X.
\ee
Thus
\begin{align}
\Omega A_\textsc{o}
&=\text{anti}(A_{1})\otimes \omega - \text{sym}(A_{\omega}) \otimes \openone_2 \\
\nonumber
&+ \text{anti}(A_{x})\otimes  Z - \text{anti}(A_{z})\otimes X,
\end{align}
and
\begin{align}
\text{anti}(\Omega A_\textsc{o})
\nonumber
&=\text{sym}(\text{anti}(A_{1}))\otimes \omega - \text{anti}(\text{sym}(A_{\omega})) \otimes \openone_2 \\
\nonumber
&+ \text{anti}(\text{anti}(A_{x}))\otimes  Z - \text{anti}(\text{anti}(A_{z}))\otimes X\\
&=\text{anti}(A_{x})\otimes  Z - \text{anti}(A_{z})\otimes X.
\end{align}
where we have used that $\text{sym}$ and $\text{anti}$ are orthogonal projectors. Finally multiplying by  $\Omega^{-1}$ on the left undoes the transformation \eqref{2by2cycle} yielding,
\begin{align}
A_\textsc{oa} =& \Omega^{-1}\text{anti}(\Omega A_\textsc{o})\\
&=\text{anti}(A_{x})\otimes  X + \text{anti}(A_{z})\otimes Z.
\end{align} 
Thus the orthogonal active part of $A$ is the sum of two $N$ by $N$ antisymmetric matrices tensored with $X$ and $Z$.

A similar analysis can be performed on the other parts of $A$, yielding
\begin{align}\label{AnaDef}
A_\textsc{oa} &= A_{x,\text{anti}} \otimes X + A_{z,\text{anti}} \otimes Z,\\
A_\textsc{op} &= A_{1,\text{anti}} \otimes \openone_{2} + A_{w,\text{sym}} \otimes \omega,\\
A_\textsc{na} &= A_{1,\text{sym}} \otimes \openone_{2} + A_{w,\text{anti}} \otimes \omega,\\
A_\textsc{np} &= A_{x,\text{sym}} \otimes X + A_{z,\text{sym}} \otimes Z,
\end{align}
where $A_{\mu,\text{sym}}$ and $A_{\mu,\text{anti}}$ are some symmetric and antisymmetric $N$ by $N$ matrices for $\mu\in\{1,w,x,z\}$.

Finally each of these can be further subdivided into its single and multi-mode parts by isolating their block diagonal elements. Note that in the expansion given by \eqref{expansiona} the block diagonal elements of, for instance, the $A_x\otimes X$ term correspond to the diagonal elements of $A_x$. Defining $A_\mu^D$ to be the diagonal elements of $A_\mu$ we find the single mode parts of each term to be,
\begin{align}\label{afactor}
A_\textsc{oa}^\textsc{s} = &A_{x,anti}^{D} \otimes X + A_{z,anti}^{D} \otimes Z = 0,\\
\nonumber
A_\textsc{op}^\textsc{s} =& A_{1,anti} ^{D} \otimes \openone_{2} + A_{w,sym}^{D} \otimes \omega= A_{w,sym}^{D} \otimes \omega,\\
\nonumber
A_\textsc{na}^\textsc{s} = &A_{1,sym}^{D} \otimes \openone_{2} + A_{w,anti}^{D} \otimes \omega = A_{1,sym}^{D} \otimes \openone_{2},\\
\nonumber
A_\textsc{np}^\textsc{s}= &A_{x,sym}^{D} \otimes X + A_{z,sym} ^{D}\otimes Z.
\end{align}
Note that the single-mode orthogonal active state-dependent part of the dynamics ($A_\textsc{oa}^\textsc{s}$) vanishes since the diagonals of an antisymmetric matrices are zero. The multi-mode  parts of each term is given by the difference between the terms and their single modes parts, $A^\textsc{m}=A-A^\textsc{s}$.

Now that we have fully partitioned the state-dependent part of our dynamics, $A$, we turn our attention to the state-independent part, $C$. As we did for $A$, we expand $C$ over the $2 \times 2$ basis \eqref{2by2basis} as 
\begin{align}\label{expansionc}
C\,=\, & \,C_1 \otimes\openone_2 + C_w \otimes \omega + C_x \otimes X + C_z \otimes Z,
\end{align}
where $C_\mu$ are $N$ by $N$ matrices for $\mu \in \lbrace 1, \omega, x, z \rbrace $. Recall that $C$ must be an antisymmetric matrix. This implies that its coefficient matrices must be either symmetric or antisymmetric depending on their accompanying tensor factor. Specifically, since $\openone_{2}$, $X$, and $Z$ are symmetric $C_1$, $C_x$, and $C_z$ must be antisymmetric. Similarly since $\omega$ is antisymmetric $C_w$ must be symmetric. 

To  begin we note that, as discussed above, $C$ is entirely non-orthogonal, $C_\textsc{n}=C$ and $C_\textsc{o}=0$. Next we will partition $C$ into its single-mode and multi-mode parts. As before this means splitting $C$ into its block-diagonal  and block-off-diagonal  elements. This again corresponds to isolating the diagonal elements of $C$'s coefficient matrices     
\begin{align}\nonumber
C_\textsc{n}^\textsc{s} =\, & \,C_1^{D}\otimes\openone_2 + C_{\omega}^{D} \otimes \omega + C_x^{D} \otimes X + C_z^{D} \otimes Z\\
=\, & \, C_{\omega}^{D} \otimes \omega,
\end{align}
where we have again exploited the fact that the diagonals of antisymmetric matrices vanish. We can identify the rest of $C$ as its multi-mode part
\begin{align}
C_\textsc{n}^\textsc{m} = C - C^\textsc{s}_\textsc{n}.
\end{align}

Next we will divide $C$ into its active and passive parts according to how it affects a system's average excitation number. Recalling equation \eqref{Activity} we can see that $C$ contributes to the average excitation number through the trace of $\Omega C$. 

As we will now argue, the multi-mode part of $C$ is entirely passive. To see why recall that by definition the multi-mode part of $C$ is block-off-diagonal. Note that multiplying by $\Omega=\openone_N\otimes\omega$ acts trivially on the first tensor factor such that $\Omega C_\textsc{n}^\textsc{m}$ is also block-off-diagonal. Thus $\Omega C_\textsc{n}^\textsc{m}$ is traceless and therefore $C_\textsc{n}^\textsc{m}$ is passive. Thus the active part of $C$ is entirely single-mode. 

Now we can compute $C$'s contribution to the change of particle number as  
\begin{align*}
\text{Tr}(\Omega \, C_\textsc{n})
&=\text{Tr}(\Omega C_\textsc{n}^\textsc{s})\\
&=\text{Tr}\Big( \openone_N\otimes \omega \ \ C_{\omega}^D \otimes \omega\Big)\\
&=-\text{Tr}\Big(C_{\omega}^D \otimes \openone_2\Big)
\end{align*}
and so the diagonal elements of $C_{\omega}$ determine if $C$ is active or not. Thus we are lead to identify the active and passive parts of $C$ as,
\begin{align}
C_\textsc{na} = C_{\omega}^{D} \otimes \omega,\\
C_\textsc{np} = C-C_\textsc{na},
\end{align}
respectively. Coincidentally these are the same terms we found when dividing $C$ into its single and multi-mode parts, $C_\textsc{n}^\textsc{s}=C_\textsc{na}=C^\textsc{s}_\textsc{na}$ and $C_\textsc{n}^\textsc{m}=C_\textsc{np}=C_\textsc{np}^\textsc{m}$.

The results of this partition are summarized in Table \ref{SUAPTable}. Note that the partition has revealed that only 9 of the potential 16 types of dynamics are realized.

\begin{table}[h]
\begin{tabular}{|r|l|l|l|l|} 
\hline
& \multicolumn{2}{c|}{\bf Active} & \multicolumn{2}{c|}{\bf Passive}\\ 
\hline
{\bf Orthogonal} 
& \quad $A_\textsc{oa}^{( \ /\textsc{m})}$ \quad 
& \quad \quad 
& \quad $A_\textsc{op}^{(\textsc{s}/\textsc{m})}$ \quad 
&  \quad \quad \\
\hline
{ \bf Non-orthogonal} 
& \quad $A_\textsc{na}^{(\textsc{s}/\textsc{m})}$ \quad 
& \quad $C_\textsc{na}^{(\textsc{s}/ \ )}$ \quad
& \quad $A_\textsc{np}^{(\textsc{s}/\textsc{m})}$ \quad 
& \quad $C_\textsc{np}^{( \ /\textsc{m})}$ \quad \\ 
\hline
\quad  
& \quad \bf S.D. \quad 
& \quad \bf S.I. \quad
& \quad \bf S.D. \quad 
& \quad \bf S.I. \quad \\ 
\hline
\end{tabular}
\caption{The results of the partition performed in Section \ref{partition}. Note each cell is divided horizontally into a state-dependent (S.D.) and state-independent (S.I.) part. The superscripts on each term indicate whether or not such terms can be single-mode (S) or multi-mode (M) or both.  An empty cell indicates the dynamics is not possible. Note that the partition has revealed that only 9 of the potential 16 types of dynamics are realized.}
\label{SUAPTable}
\end{table}

\subsection{Complete Positivity}
While the above partition produced nine distinct types of dynamics, not all of these are completely positive in isolation. As we will see, in order to be completely positive any non-orthogonal dynamics (either $C\neq0$ or $A_\textsc{n}\neq0$) must be accompanied by a non-zero amount of noise ($A_\textsc{na}^\textsc{s}\neq0$). 

We prove this by showing that for completely postive dynamics $C\neq0$ implies $A_\textsc{n}\neq0$ which itself implies $\text{Tr}(\text{A}_\textsc{n})<0$. Following this we will show that $A_\textsc{na}^\textsc{s}$ is the only part of the dynamics which contributes to this trace. Later, in Section \ref{Naming}, we will show why it is appropriate to interpret $A_\textsc{na}^\textsc{s}$ as generating noise. 

To begin our proof, we first write the completely positive condition \eqref{diffcptpcond} in terms of the partition as 
\bel{DiffCpCondPartitioned}
2A_\textsc{n}+\ii C\leq0.
\ee
If $A_\textsc{n}=0$, this reduces to $\ii C\leq0$. Taking the complex conjugate of this equation we find that $-\ii C\leq0$ or equivalently $\ii C\geq0$, where we recall that $C$ is a real matrix. The only way that both of these inequalities can be true is if $C=0$. Taking the contrapositive of this result we find that $C \neq 0$ implies $A_\textsc{n}\neq0$.

By adding \eqref{DiffCpCondPartitioned} to its complex conjugate we see that completely positive dynamics has $A_\textsc{n}\leq0$. This means that the eigenvalues of $A_\textsc{n}$ are all real and non-positive. From this it immediately follows that their sum is non-positive, $\text{Tr}(A_\textsc{n})\leq0$. Since the eigenvalues are non-positive, their sum can only vanish if all of the eigenvalues are themselves zero. In other words, $\text{Tr}(A_\textsc{n})=0$ implies $A_\textsc{n}=0$. Thus if $A_\textsc{n}$ does not vanish then neither can $\text{Tr}(A_\textsc{n})$ such that we have $\text{Tr}(A_\textsc{n})<0$.

As we have now seen, in order for the dynamics to be completely positive, the presence of any non-orthogonal dynamics implies that $\text{Tr}(A_\textsc{n})<0$. Using the partition described above we will now identify which parts of the dynamics contributes to this trace. 

To begin we compute $\text{Tr}(A_{\textsc{np}})$, that is the part of this trace coming from passive dynamics. From \eqref{afactor} we have
\begin{align}
\text{Tr}(A_\textsc{np}) &= \text{Tr}(A_{x,\text{sym}} \otimes X + A_{z,\text{sym}} \otimes Z)\\
\nonumber
&= \text{Tr}(A_{x,\text{sym}}) \ \text{Tr}(X) + \text{Tr}(A_{z,\text{sym}}) \  \text{Tr}(Z)\\
\nonumber
&= 0
\end{align}
and so the dynamics that contribute to $\text{Tr}(A_\textsc{n})$ must be active. Next we can argue that since the multi-mode parts of $A$ are block-off-diagonal  they cannot contribute to this trace either. Thus the dynamics  contributing to $\text{Tr}(A_\textsc{n})$ must be single-mode. Thus the only part of the dynamics contributing to $\text{Tr}(A_\textsc{n})$ is $A_\textsc{na}^\textsc{s}$.

Hence, completely positive non-orthogonal dynamics must have $A_\textsc{na}^\textsc{s}\neq0$. As we will see in the next section, this type of dynamics can be interpreted as generating noise. The fact that non-orthogonal (and more generally non-unitary) dynamics must be noisy is well known and holds outside of the Gaussian context we are discussing here. The novel connection here is that for Gaussian fermionic systems this noise must be active. That is particle-number non-conserving and requiring an environment for its particle exchange. Though some active dynamics do not necessarily require an environment to exchange particles with, we also comment that such dynamics are not Gaussian, that is they do not map Guassian states to Gaussian states and thus cannot be regarded as fermionic Gaussian maps.

Any non-trivial interaction with an environment must involve particle/excitation exchange with that environment, for at least some initial states. In other words, any completely positive interaction with an environment having no particle/excitation exchange is orthogonal and thus can be implemented/explained/modelled without that environment. 

Moreover, since any state-independent dynamics is necessarily non-orthogonal it must also be noisy ($A_\textsc{na}^\textsc{s}\neq0$) in order to be completely positive. Since this noise term is state dependent, all completely positive fermionic Gaussian dynamics must include a state-dependent part.

Lastly, it is worth noting that complete  positivity can be violated if the dynamics is non-Markovian \cite{nonMarkovian}. In such cases non-orthogonal dynamics could in principle appear without an additional noise term.

\section{Understanding the Different Components}\label{Naming}

Having completed the classification and partition of the dynamics, we now study the different  types of dynamics  that are possible for fermionic systems and how they relate to the partition we performed above. Note that it is sufficient to consider systems composed of one or two modes ($N=1,2$) in order to build illustrative examples of every type of dynamics. We summarize the results of this section in Table \ref{Table2}.

\subsection{Single Mode Dynamics}

As discussed above, taking $N=1$ greatly simplifies the state space available to a fermionic system. Specifically, all physical single-mode states are thermal states with respect to their free Hamiltonian \eqref{FermionicFreeHam}. Recall these states have a covariance matrix given by \eqref{gamform},
\begin{align}\label{gcov}
\Gamma = \nu \, \omega; \qquad \omega=\begin{pmatrix}
0 & 1 \\
-1 & 0
\end{pmatrix},
\end{align}
where $\nu=\text{tanh}(\beta E/2)$ is a temperature monotone and $E$ is the mode's excitation energy. Recall the parameter $\nu$ is related to the expected excitation number of the mode as $\langle\hat{n}\rangle=\frac{1}{2}-\frac{\nu}{2}$. In order for  a state with covariance matrix \eqref{gcov} to be a physically valid state we must have $-1\leq\nu\leq1$ such that $0\leq\langle\hat{n}\rangle\leq1$.

The dynamics of a single mode are equally trivial. Since $N=1$ the coefficient matrices of $A$ and $C$ in \eqref{expansiona} and \eqref{expansionc} are just scalars. Specifically,  
\be
A_{N=1}= a_\textsc{op} \ \omega
-a_\textsc{na} \ \openone_2
+a_\textsc{np,x} \ X
+a_\textsc{np,z} \ Z,
\ee
for some real parameters $a_\textsc{op}$, $a_\textsc{na}$, $a_\textsc{np,x}$,  and $a_\textsc{np,z}$. Likewise,
\be
C_{N=1}= c_\textsc{na} \omega,
\ee
for some real parameter $c_\textsc{na}$. The complete positivity condition \eqref{cptpcondition} here reduces to
\be
a_\textsc{na}\geq\sqrt{a_\textsc{np,x}^2+a_\textsc{np,z}^2+c_\textsc{na}^2/4} \geq 0. 
\ee
Note that, as  discussed above, the presence of any non-orthogonal dynamics necessitates the presence of $a_\textsc{na}\neq0$.

\subsubsection{Orthogonal Single-Mode Dynamics}
According to the partition described above, the only type of dynamics that is orthogonal (i.e., unitary in Hilbert space) and single-mode is also passive and state-dependent. Such dynamics is generated by $A_\textsc{op}^\textsc{s}$. For one  mode ($N=1$) the generator of this dynamics is of the form
\be
A_\textsc{op}^\textsc{s}= -E \, \omega
\ee
for some real parameter $E$. 

To interpret this type of dynamics, we   compute its effective Hamiltonian. From \eqref{Heff} we find 
\begin{align}
\hat{H}_\text{eff}
&=\frac{-\ii}{2}E(\hat{x}_1\hat{p}_1-\hat{p}_1\hat{x}_1)
=E(\hat{n}_1-\frac{1}{2}).
\end{align}
Thus we can interpret this dynamics as the free evolution of the mode, where $E$ is its excitation energy.

From \eqref{ode} we  compute the effect of this dynamics on the system's covariance matrix, finding 
\begin{align}
\Gamma'(t)
&=A_\textsc{op}^\textsc{s} \ \Gamma(t)
+ \Gamma(t) \ (A_\textsc{op}^\textsc{s}{})^\intercal\\
&=-E \, \nu(t) \,  (\omega \omega+\omega \omega^\intercal)\\
&=0
\end{align}
that is, the dynamics that does not change the state of the system. One may have anticipated this by recalling that for one  mode ($N=1$) all physical states are thermal and therefore stationary under free evolution. This could also have been anticipated by noting that this dynamics is passive, and so cannot change $\langle\hat{n}\rangle$ (and therefore cannot change $\nu$ or $\Gamma$).

\subsubsection{Non-Orthogonal Single-Mode Dynamics}

The partition described above identifies three types of non-orthogonal single-mode dynamics ($A_\textsc{na}^\textsc{s}$, $C_\textsc{na}^\textsc{s}$, and $A_\textsc{np}^\textsc{s}$) and which we will now discuss in turn. 

As discussed above, complete  positivity requires that any non-orthogonal dynamics is accompanied by some $A_\textsc{na}^\textsc{s}\neq0$. For one  mode ($N=1$) the generator of this dynamics is of the form
\be
A_\textsc{na}^\textsc{s}= - r \, \openone_2
\ee
for some real parameter $r$. Complete positivity requires $r\geq 0$. Using equation \eqref{LindbladOp} we can associate this dynamics with Lindblad operators that are linear in the Majorana operators.
\be
\hat{L}_1=\sqrt{2\,r} \ \hat{x}_1,
\qquad
\hat{L}_2=\sqrt{2\,r} \ \hat{p}_1.
\ee

Next we compute the effect of this dynamics   on the system's covariance matrix, finding
\be
\nu'(t)=-2 r \, \nu(t).
\ee
Thus this dynamics causes $\nu$ to decay exponentially to zero at a rate $2\, r$. Once $\nu=0$ the state is maximally mixed. Thus we can identify $A_\textsc{na}^\textsc{s}$ as adding noise to the system.

Next let  us now look at state-independent active non-orthogonal single-mode dynamics, that is $C_\textsc{na}^\textsc{s}$. This type of dynamics is generated by
\be
C_\textsc{na}^\textsc{s}= c \, \omega
\ee
for some real parameter $c$. In order to be completely positive this dynamics must be accompanied by a minimum level of noise. Specifically, $A_\textsc{na}^\textsc{s}= - r \, \openone_2$ with $r\geq\vert c/2\vert$. Using equation \eqref{LindbladOp} we can associate this dynamics with Lindblad operators that are linear in the Majorana operators.
\be
\hat{L}_1=\sqrt{r-c/2} \ \hat{a}_1,
\qquad
\hat{L}_2=\sqrt{r+c/2} \ \hat{a}_1^\dagger.
\ee

Computing the effect of this dynamics we find
\be
\nu'(t)=-2 r \  \nu(t)+c.
\ee
This results in $\nu$ being exponentially attracted towards $\nu(\infty)=c/2 r$ at a rate $2\,r$. Note that the complete positivity of the dynamics implies that this final state of the system is physical, i.e. $-1\leq\nu(\infty)\leq1$. In the limiting case where $c=\pm 2r$ the system's final state has $\nu(\infty)=\pm 1$. These are the system's two pure states, $\ket{0}$ and $\ket{1}$. Hence we identify $C_\textsc{na}^\textsc{s}$ dynamics as purifying the state. 

Finally, let us look at state-dependent passive non-orthogonal single-mode dynamics, that is $A_\textsc{np}^\textsc{s}$. This type of dynamics is generated by
\begin{align}\label{nonorth}
A_\textsc{np}^\textsc{s}= b_x \, X +b_z \, Z
\end{align}
for some real parameters $b_x$ and $b_z$. In order to be completely positive this dynamics must be accompanied by a minimum level of noise. Specifically, $A_\textsc{na}^\textsc{s}= - r \, \openone_2$ with $r\geq\sqrt{b_x^2+b_z^2}$.

Note that as we discussed in Section \ref{SingleModeMultiMode} our classification scheme is invariant under a change of local basis. Thus without loss of generality, it is sufficient to only investigate the $b_x$ term. Using equation \eqref{LindbladOp} we can associate this term with Lindblad operators that are linear in the Majorana operators.
\be
\hat{L}_1=\sqrt{r-b_x} \ (\hat{x}_1+\hat{p}_1),
\qquad
\hat{L}_2=\sqrt{r+b_x} \ (\hat{x}_1-\hat{p}_1).
\ee

As we saw with free evolution, this dynamics cannot affect $\nu$ since it is passive. However, this does not mean that this dynamics is completely trivial. As we will see in the next section, this dynamics affects the evolution of the mode's correlations with other uncoupled systems.

\subsection{Multi-mode Dynamics}
A generic covariance matrix for $N=2$ modes can be written as,
\begin{align}\label{2modecm}
\Gamma = \begin{pmatrix}
0 & \nu_1 & g_1 & g_2\\
-\nu_1 & 0 & g_3 & g_4\\
-g_1 & -g_3 & 0 & \nu_2\\
-g_2 & -g_4 & -\nu_2 & 0
\end{pmatrix}
\end{align}
for some local temperature monotones $\nu_1$ and $\nu_2$ and four correlation numbers: $g_1$,  $g_2$,  $g_3$, and $g_4$. Multi-mode dynamics couples these parameters together via the master equation \eqref{ode}. Specifically, the six parameters of the covariance matrix, \mbox{$\bm{g}=\{\nu_1,\nu_2,g_1,g_2,g_3,g_4\}^\intercal$}, will evolve under a system of first order differential equations as, 
\be
\bm{g}'(t)=\mathcal{A} \,  \bm{g}(t)+\mathcal{C}
\ee
for some $6$ by $6$ real-valued matrix, $\mathcal{A}$, and $6$ dimensional real-valued vector, $\mathcal{C}$.

In order to examine the effect of multi-mode dynamics we will convert the dynamics into the above form and then perform an eigen-decomposition of $\mathcal{A}$.

\subsubsection{Revisiting Single-Mode Dynamics}
Before we look at multi-mode dynamics let us look at how single-mode dynamics affect existing correlations. 

First we will look at the effect of free rotation on the system's correlations. Taking each mode to have excitation energies, $E_1$ and $E_2$, their free evolution is generated by
\begin{align}\label{aparameter}
A=-E_1 \, \omega \oplus -E_2 \omega.
\end{align}
Computing from \eqref{ode} the rate of change of the covariance matrix  using \eqref{aparameter} and \eqref{2modecm} we find 
 \begin{align}
\frac{\d\Gamma}{\d t} 
= \begin{pmatrix}
0 & 0 & -E_2 \, g_2 - E_1 \, g_3 & E_2 \, g_1 - E_1 \, g_4\\
  & 0 & E_1 \, g_1 - E_2 \, g_4 & E_1 \, g_2 + E_2 \, g_3\\
& & 0 & 0\\
& & & 0
\end{pmatrix}
\end{align}
where the lower left triangle is the negation of the upper right one. Note that as expected the free rotation does not affect the reduced state of either system; $\nu_1$ and $\nu_2$ are constant. From this we can read off $\mathcal{A}$ as
\begin{align}
\mathcal{A}=
\begin{pmatrix}
0 & 0 & 0 & 0 & 0 & 0\\
0 & 0 & 0 & 0 & 0 & 0\\
0 & 0 & 0 & -E_2 & -E_1 & 0\\
0 & 0 & E_2 & 0 & 0 & -E_1\\
0 & 0 & E_1 & 0 & 0 & -E_2\\
0 & 0 & 0 & E_1 & E_2 & 0
\end{pmatrix}.
\end{align}
To analyze how the correlations effect each other we can diagonalize $\mathcal{A}$. However in this case it is more convenient to diagonalize $\mathcal{A}^2$, which is related to the second order differential equations $\bm{g}''(t)=\mathcal{A}^2\bm{g}(t)$ (note $\mathcal{C}=0$). The result is
\begin{align}
\frac{\d^2}{\d t^2}
\begin{pmatrix}
\nu_1\\
\nu_2\\
g_1+g_4\\
g_1-g_4\\
g_2+g_3\\
g_2-g_3
\end{pmatrix}
=\text{diag}
\begin{pmatrix}
0\\
0\\
-(E_1-E_2)^2\\
-(E_1+E_2)^2\\
-(E_1+E_2)^2\\
-(E_1-E_2)^2
\end{pmatrix}
\begin{pmatrix}
\nu_1\\
\nu_2\\
g_1+g_4\\
g_1-g_4\\
g_2+g_3\\
g_2-g_3
\end{pmatrix} 
\end{align}
where \text{diag} is the usual notation for a diagonal matrix with its non-zero elements given by the argument. Thus the correlations to rotate among themselves. In particular the $g_1+g_4$ and $g_2-g_3$ correlations oscillate at a rate $E_1-E_2$ and the $g_1-g_4$ and $g_2+g_3$ correlations oscillate at a rate $E_1+E_2$.

Next, let us examine the effect of $A_\textsc{np}^\textsc{s}$ on multi-mode correlations. Since this dynamics is non-orthogonal we must introduce a certain amount of noise to make it completely positive. Restricting our attention to the $b_x$ term in \eqref{nonorth} we can take
\begin{align*}
A = (-r \, \openone_2+b_x \, X)\oplus  {0}_2,
\end{align*}
where $0_2$ is the 2 by 2 zero matrix and $r\geq\vert b_x \vert$ is required for complete positivity.

Computing $\mathcal{A}$ and diagonalizing it we find
\begin{align}
\frac{\d}{\d t}
\begin{pmatrix}
\nu_1\\
\nu_2\\
g_1+g_3\\
g_1-g_3\\
g_2+g_4\\
g_2-g_4\\
\end{pmatrix}
=\text{diag}
\begin{pmatrix}
-2\,r \\
0\\
-(r-b_x)\\
-(r+b_x)\\
-(r-b_x)\\
-(r+b_x)
\end{pmatrix}
\begin{pmatrix}
\nu_1\\
\nu_2\\
g_1+g_3\\
g_1-g_3\\
g_2+g_4\\
g_2-g_4\\
\end{pmatrix}
\end{align}
such that unless $b_x=\pm r$, all of the parameters of the covariance matrix (except $\nu_2$) are driven to zero. That is, eventually the first mode becomes maximally mixed and all of its correlations with the second mode are broken. The effect of $A_\textsc{np}^\textsc{s}$ is to modify the rates at which the parameters decay. In the limiting case where $b_x=\pm r$ then the $g_1\pm g_3$ and $g_2\pm g_4$ correlations are completely shielded from this decay. For the purpose of our classification we will call this dynamics `correlation shielding'.

Repeating this analysis on the $b_z$ term we find,
\begin{align}\label{NPanal}
\frac{\d}{\d t}
\begin{pmatrix}
\nu_1\\
\nu_2\\
g_1\\
g_2\\
g_3\\
g_4
\end{pmatrix}
=\text{diag}
\begin{pmatrix}
-2\,r \\
0\\
-(r-b_z)\\
-(r-b_z)\\
-(r+b_z)\\
-(r+b_z)
\end{pmatrix}
\begin{pmatrix}
\nu_1\\
\nu_2\\
g_1\\
g_2\\
g_3\\
g_4
\end{pmatrix}
\end{align}
where $r \geq |b_z|$ is required for complete positivity. Note that as before all of the parameters of the covariance matrix (except $\nu_2$) are again driven to zero unless $ b_z = \pm r$.  If $b_z=r$ then the $g_1$ and $g_2$ correlations are shielded from decay, and if $b_z=-r$ the $g_3$ and $g_4$ correlations are shielded.

\subsubsection{Orthogonal, passive and state-dependent dynamics}
For $N=2$ the multi-mode orthogonal passive state-dependent dynamics are given by
\begin{align}
A_\textsc{op} = \begin{pmatrix} 0_2&  a_1\openone_2 + a_w \omega  \\ -a_1 \openone_2  + a_w \omega & 0_2 \end{pmatrix}.
\end{align}
Since this dynamics is orthogonal we can compute its effective Hamiltonian from equation \eqref{Heff}, obtaining
\be
\hat{H}_\text{eff}
=\frac{\ii}{2} b_1 (\hat{x}_1\hat{x}_2+\hat{p}_1\hat{p}_2)
+\frac{\ii}{2} b_w (\hat{x}_1\hat{p}_2-\hat{p}_1\hat{x}_2)
+\text{h.c.}
\ee
Written in terms of the modes' creation and annihilation operators this is
\be
\hat{H}_\text{eff}
=(b_w+\ii b_1)\hat{a}_1\hat{a}_2^\dagger
-(b_w-\ii b_1)\hat{a}_1^\dagger\hat{a}_2.
\ee
Note that every term in this effective Hamiltonian has an equal number of creation and annihilation operators, such that it is manifestly number conserving/passive. We should also note that these are the type of terms that would arise from a ``rotating wave''-like approximation.

To analyze the effect of this dynamics let us restrict our attention to the $b_w$ term. Computing and diagonalizing $\mathcal{A}^2$ we find,
\begin{align}
\frac{\d^2}{\d t^2}
\begin{pmatrix}
\nu_1-\nu_2\\
\nu_1+\nu_2\\
g_1+g_4\\
g_1-g_4\\
g_2+g_3\\
g_2-g_3
\end{pmatrix}
=\text{diag}
\begin{pmatrix}
-4 \, b_w^2 \\
0\\
-4 \, b_w^2\\
0\\
0\\
0
\end{pmatrix}
\begin{pmatrix}
\nu_1-\nu_2\\
\nu_1+\nu_2\\
g_1+g_4\\
g_1-g_4\\
g_2+g_3\\
g_2-g_3
\end{pmatrix}.
\end{align}
Thus we can see that this dynamics causes the difference in the modes' excitation level, $\nu_1-\nu_2$, and the $g_4+g_1$ correlations to oscillate at a rate $2 \, b_w$. 

Note that the remaining variables  do not grow linearly with time but are constant. This can be shown by considering $\mathcal{A}$ (instead of $\mathcal{A}^2$), for which the equations reduce to
\begin{align}
\frac{\d}{\d t}
\begin{pmatrix}
\nu_1-\nu_2\\
g_1+g_4
\end{pmatrix}
=\begin{pmatrix}
0 & 2 \, b_w\\
-2 \, b_w & 0
\end{pmatrix}
\begin{pmatrix}
\nu_1-\nu_2\\
g_1+g_4
\end{pmatrix}
\end{align}
with all other first derivatives vanishing.

Repeating our analysis on the $b_1$ term we find the same result as above but the $g_2-g_3$ correlation oscillates instead.

\subsubsection{Orthogonal active and state-dependent dynamics}
For $N=2$ the multi-mode orthogonal active state-dependent dynamics are given by 
\begin{align}
A_\textsc{oa} 
=\begin{pmatrix} 
0_2&  b_x X+ b_z Z  \\ 
-b_x X - b_z Z & 0_2 \end{pmatrix}.
\end{align}
Since this dynamics is orthogonal we can again from equation \eqref{Heff} compute its effective Hamiltonian, obtaining 
\be
\hat{H}_\text{eff}
=\frac{\ii}{2} b_x (\hat{x}_1\hat{p}_2+\hat{p}_1\hat{x}_2)
+\frac{\ii}{2} b_z (\hat{x}_1\hat{x}_2-\hat{p}_1\hat{p}_2)
+\text{h.c.}
\ee
Written in terms of creation and annihilation operators this is
\be
\hat{H}_\text{eff}
=(b_x+\ii b_z)\hat{a}_1^\dagger\hat{a}_2^\dagger
-(b_x-\ii b_z)\hat{a}_1\hat{a}_2.
\ee
Note that every term in this effective Hamiltonian has an unequal number of creation and annihilation operators, such that it is manifestly number non-conserving/active. We should also note that these are the terms which would be dropped when taking the ``rotating wave''-like approximation.

To analyze the effect of this dynamics, let us restrict our attention to the $b_x$ term. Computing and diagonalizing $\mathcal{A}^2$ we find
\begin{align}
\frac{\d^2}{\d t^2}
\begin{pmatrix}
\nu_1+\nu_2\\
\nu_1-\nu_2\\
g_1+g_4\\
g_1-g_4\\
g_2+g_3\\
g_2-g_3
\end{pmatrix}
=\text{diag}
\begin{pmatrix}
-4 b_x^2 \\
0\\
0\\
-4 b_x^2\\
0\\
0
\end{pmatrix}
\begin{pmatrix}
\nu_1+\nu_2\\
\nu_1-\nu_2\\
g_1+g_4\\
g_1-g_4\\
g_2+g_3\\
g_2-g_3
\end{pmatrix}.
\end{align}
Thus we can see this dynamics causes the total excitation level, $\nu_1+\nu_2$, and the $g_1-g_4$ correlations to oscillate at a rate $2 b_x$. One can imagine the modes both becoming more excited and unexcited in unison while correlations between them rise and fall. As before, the remaining variables do not grow linearly with time but are constant.

Repeating our analysis on the $b_z$ term we find the same result as above but the $g_2+g_3$ correlations oscillate instead.

\subsubsection{Non-orthogonal active and state-dependent dynamics}

For $N=2$ the multi-mode non-orthogonal active state-dependent dynamics are given by,
\begin{align}
A_\textsc{na} = \begin{pmatrix} 
0_2&  b_1\openone_2 + b_w \omega  \\ 
b_1 \openone_2  - b_w \omega & 0_2 \end{pmatrix}.
\end{align}
Since this dynamics is non-orthogonal we must introduce a certain amount of noise to make it completely positive. Restricting our attention to the $b_{w}$ term we first examine,
\begin{align}
A
= \begin{pmatrix} -r\, \openone_2 &  b_w \,  \omega  \\ 
- b_w \, \omega & -r\, \openone_2 \end{pmatrix},
\end{align}
where $r\geq\vert b_w\vert$ is required for complete positivity.

Using equation \eqref{LindbladOp} we can associate this term with Lindblad operators that are linear in the Majorana operators.
\begin{align}
\hat{L}_1&=\sqrt{r-b_w} \ (\hat{x}_1+\hat{p}_2),
\quad
\hat{L}_2=\sqrt{r-b_w} \ (\hat{p}_1-\hat{x}_2),\\
\hat{L}_3&=\sqrt{r+b_w} \ (\hat{x}_1-\hat{p}_2),
\quad
\hat{L}_4=\sqrt{r+b_w} \ (\hat{p}_1+\hat{x}_2).
\end{align}
 
Computing and diagonalizing $\mathcal{A}$ we find,
\begin{align}
\frac{\d}{\d t}\!
\begin{pmatrix}
\nu_1+\nu_2+g_1+g_4\\
\nu_1+\nu_2-g_1-g_4\\
\nu_1-\nu_2\\
g_1-g_4\\
g_2+g_3\\
g_2-g_3
\end{pmatrix}
\!=\!\text{diag}
\begin{pmatrix}
-2 (r+b_w) \\
-2 (r-b_w) \\
-2 r \\
-2 r\\
-2 r\\
-2 r
\end{pmatrix}
\begin{pmatrix}
\nu_1+\nu_2+g_1+g_4\\
\nu_1+\nu_2-g_1-g_4\\
\nu_1-\nu_2\\
g_1-g_4\\
g_2+g_3\\
g_2-g_3
\end{pmatrix}.
\end{align}
Note that unless $b_w=\pm r$, all of the parameters of the covariance matrix are suppressed to zero; The modes become maximally mixed and uncorrelated.  The effect of this dynamics is to modify the rates at which the parameters decay. In the limiting case where $b_w=\pm r$ the final state may still have some correlations. For example if $b_w=-r$ then the sum, $\nu_1+\nu_2+g_1+g_4$, is preserved resulting in the final state
\be
\Gamma(\infty)
=\begin{pmatrix}
0 & k & k & 0\\
-k & 0 & 0 & k\\
-k & 0 & 0 & k\\
0 &  -k & -k & 0\\
\end{pmatrix},
\ee
where $k=\frac{1}{4}(\nu_1+\nu_2+g_1+g_4)\vert_{t=0}$. 

The $b_1$ term provides similar phenomenology, shielding either the sum $\nu_1+\nu_2+g_3-g_2$ or $\nu_1+\nu_2-g_3+g_2$.

\subsubsection{Non-orthogonal, passive and state-dependent}

For $N=2$ the multi-mode, non-orthogonal, passive, state-dependent dynamics are given by,
\begin{align}
A_\textsc{np} = \begin{pmatrix} 
0_2 &  b_x \, X + b_z \, Z \\ b_x \, X  + b_z \, Z & 0_2 \end{pmatrix}.
\end{align}
Since this dynamics is non-orthogonal we must introduce a certain amount of noise to make it completely positive. Restricting our attention to the $b_x$ term we have,
\begin{align}
A = \begin{pmatrix} -r \, \openone_2 & b_x \, X  \\ 
b_x \, X & -r \, \openone_2 \end{pmatrix},
\end{align}
where $r\geq\vert b_x\vert$ is required for complete positivity.

Using equation \eqref{LindbladOp} we can associate this term with  Lindblad operators which are linear in the Majorana operators.
\begin{align}
\hat{L}_1&=\sqrt{r-b_w} \ (\hat{x}_1+\hat{p}_2),
\quad
\hat{L}_2=\sqrt{r-b_w} \ (\hat{p}_1+\hat{x}_2),\\
\hat{L}_3&=\sqrt{r+b_w} \ (\hat{x}_1-\hat{p}_2),
\quad
\hat{L}_4=\sqrt{r+b_w} \ (\hat{p}_1-\hat{x}_2).
\end{align}

Computing and diagonalizing $\mathcal{A}$ we find,
\begin{align}
\frac{\d}{\d t}\!
\begin{pmatrix}
\nu_1+\nu_2\\
\nu_1-\nu_2+g_1-g_4\\
\nu_1-\nu_2-g_1+g_4\\
g_4+g_1\\
g_2+g_3\\
g_2-g_3
\end{pmatrix}
\!=\!\text{diag}
\begin{pmatrix}
-2 r \\
-2 (r-b_x) \\
-2 (r+b_x) \\
-2 r\\
-2 r\\
-2 r
\end{pmatrix}
\begin{pmatrix}
\nu_1+\nu_2\\
\nu_1-\nu_2+g_1-g_4\\
\nu_1-\nu_2-g_1+g_4\\
g_4+g_1\\
g_2+g_3\\
g_2-g_3
\end{pmatrix}.
\end{align}
Once again, unless $b_x=\pm r$, all the parameters are suppressed to zero. The effect of this dynamics is to modify the rates at which the parameters decay. In the limiting case where $b_x=\pm r$ the final state may still have excitations and correlations. For example if $b_x=-r$ then the sum $\nu_1-\nu_2+g_1-g_4$ is preserved resuting in the state,
\be
\Gamma(\infty)
=\begin{pmatrix}
0 & k & k & 0\\
-k & 0 & 0 & -k\\
-k & 0 & 0 & -k\\
0 & k & k & 0\\
\end{pmatrix},
\ee
where $k=\frac{1}{4}(\nu_1-\nu_2+g_1-g_4)\vert_{t=0}$. 

The $b_z$ term provides similar phenomenology, shielding either the sum $\nu_1-\nu_2+g_2+g_3$ or $\nu_1-\nu_2-g_2-g_3$.

\subsubsection{Non-orthogonal, passive and state-independent dynamics}
The final type of dynamics identified by the partition described above is given by
\be
C_\textsc{np}^\textsc{s}
=\begin{pmatrix}
0 & 0 & c_1 & c_2 \\ 
0 & 0 & c_3 & c_4 \\ 
-c_1 & -c_3 & 0 & 0 \\ 
-c_2 & -c_4 & 0 & 0 \\ 
\end{pmatrix}.
\ee
This dynamics adds directly to the $g_1$, $g_2$, $g_3$, and $g_4$ correlations.

Since this dynamics is non-orthogonal we must introduce some noise to make it completely positive. Taking
\begin{align}
A = \begin{pmatrix} -r \, \openone_2 & 0\\ 
0 & -r \, \openone_2 \end{pmatrix},
\qquad
C
=\begin{pmatrix}
0 & 0 & c_1 & c_2 \\ 
0 & 0 & c_3 & c_4 \\ 
-c_1 & -c_3 & 0 & 0 \\ 
c_2 & -c_4 & 0 & 0 \\ 
\end{pmatrix}
\end{align}
we   compute $\mathcal{A}$ and $\mathcal{C}$ to find
\begin{align}
\frac{\d}{\d t}
\begin{pmatrix}
\nu_1\\
\nu_2\\
g_1\\
g_2\\
g_3\\
g_4
\end{pmatrix}
=\text{diag}
\begin{pmatrix}
-2 r \\
-2 r \\
-2 r \\
-2 r\\
-2 r\\
-2 r
\end{pmatrix}
\begin{pmatrix}
\nu_1\\
\nu_2\\
g_1\\
g_2\\
g_3\\
g_4
\end{pmatrix}
+
\begin{pmatrix}
0\\
0\\
c_1\\
c_2\\
c_3\\
c_4
\end{pmatrix}.
\end{align}
The solution to these equations have both $\nu_1$ and $\nu_2$ decaying to zero at a rate $2 r$, while the correlation $g_i$ decays to \mbox{$g_i(\infty)=c_i/2r$} at a rate $2 r$.

Note that as we discussed in Section \ref{SingleModeMultiMode} our classification scheme is invariant under a change of local basis. Using this freedom we can take a representative scenario with $c_2=0$ and $c_3=0$. In this case  we can assign equation \eqref{LindbladOp} to this type of dynamics with Lindblad operators that are linear in the Majorana operators.
\begin{align}
\hat{L}_1&=\sqrt{r-c_1/2} \ (\hat{x}_1-\ii\,\hat{x}_2),
\quad
\hat{L}_2=\sqrt{r+c_1/2} \ (\hat{x}_1+\ii\,\hat{x}_2),\\
\hat{L}_3&=\sqrt{r-c_4/2} \ (\hat{p}_1-\ii\,\hat{p}_2),
\quad
\hat{L}_4=\sqrt{r+c_4/2} \ (\hat{p}_1+\ii\,\hat{p}_2). 
\end{align}

\begin{table*}
\begin{tabular}{||cccc||M{5cm}||}
\hline 
\quad Single-mode? \quad & \quad Orthogonal? \quad  &  \quad Passive? \quad  &  \quad State-Dependent? \quad  &    Name        \\
 \quad (else Multi-mode) \quad  &  \quad (else Non-orthogonal) \quad  &  \quad (else Active) \quad  &  \quad (else Independent) \quad  & of dynamics\\
\hline Yes &        Yes &        Yes &        Yes & $A_\textsc{op}^\textsc{s}$: \ Free Evolution\\
\hline Yes &        Yes &        Yes &         No &        Not Possible \\
\hline Yes &        Yes &         No &        Yes & Not Possible \\
\hline Yes &        Yes &         No &         No & Not Possible \\
\hline Yes &         No &        Yes &        Yes & $A_\textsc{np}^\textsc{s}$:       Correlation Shielding \\
\hline Yes &         No &        Yes &         No & Not Possible \\
\hline Yes &         No &         No &        Yes & $A_\textsc{na}^\textsc{s}$: Noise \\
\hline Yes &         No &         No &         No & $C_\textsc{na}^\textsc{s}$: Purifying \\
 \hline No &        Yes &        Yes &        Yes & $A_\textsc{op}^\textsc{m}$: Multi-mode Rotation\\
 \hline No &        Yes &        Yes &         No &        Not Possible \\
 \hline No &        Yes &         No &        Yes & $A_\textsc{oa}^\textsc{m}$: Multi-mode Counter Rotation \\
 \hline No &        Yes &         No &         No &        Not Possible \\
 \hline No &         No &        Yes &        Yes & $A_\textsc{na}^\textsc{m}$:  Multi-mode Active Corr. Shielding  \\
 \hline No &         No &        Yes &         No & $C_\textsc{np}^\textsc{m}$: Correlating \\
 \hline No &         No &         No &        Yes & $A_\textsc{np}^\textsc{m}$:   Multi-mode Passive Corr. Shielding  \\
 \hline No &         No &         No &         No &        Not Possible \\
\hline
\end{tabular}
\caption{ The partition performed in Sec. \ref{partition} results in nine distinct types of open fermionic Gaussian dynamics. Examples of each  (and justifications for their names) are presented in Sec. \ref{Naming}. To summarize, in order to be completely positive, any non-orthogonal dynamics must include some noise, which tends to break any existing correlations and drives the state towards being maximally mixed. The ``correlation shielding'' dynamics slows (and in extreme cases stops) the decay of certain types of correlations with third parties. The ``purifying'' dynamics prevents the noise from making the system maximally mixed. The ``multi-mode correlation shielding'' dynamics preserve various correlations between different modes. The ``multimode rotation'' dynamics allow excitations to be transferred between two modes. Finally, the ``multimode counter rotation'' dynamics has two modes excite and de-excite in unison.    
}
\label{Table2}
\end{table*}

\section{Comparison with Bosonic Gaussian Dynamics}

The mathematical structures underlying bosonic and fermionic GQM are very similar, but lead to vastly different phenomenology. Additional comparisons of bosonic and fermionic Gaussian systems can be found in \cite{FBComp1} and \cite{FBComp2}.

Fundamentally their differences begin with how their (anti-)commutation relations are described on the system's phase space. In the fermionic/bosonic case we have
\be
\{\hat{r}_n,\hat{r}_m\}=\delta_{nm}\hat{\openone}
\quad\text{vs.}\quad
[\hat{r}_n,\hat{r}_m]=\Omega_{nm}\hat{\openone}.
\ee
For fermionic systems symmetric combinations of Majorana operators are associated with the identity matrix on phase space whereas for bosonic systems antisymmetric combinations of quadrature operators are associated with the symplectic matrix $\Omega$.

In either case, Gaussian states are fully described by the system's first and second moments. In the fermionic case, non-trivial linear combinations of the Majorana operators are unphysical so all  first moments vanish. Moreover the symmetric part of the second moments are fixed by the commutation relations. Thus all that is left is the antisymmetric covariance matrix $\Gamma_{nm}=\langle\ii[\hat{r}_n,\hat{r}_m]\rangle$. In the bosonic case, the system may have non-trivial first moments (allowing for displaced/coherent states) and the symmetric part of the system's second moments are non-trivial, that is the system's covariance matrix $\sigma_{nm}=\langle{\hat{r}_n,\hat{r}_m}\rangle$. The overall difference is that fermionic Gaussian states are more restricted then bosonic ones.

In either case, the complete positivity condition is stated as the following matrix inequality for both bosonic and fermionic sytems:
\be
-\openone_{2N}\leq\ii\Gamma\leq\openone_{2N}
\quad\text{vs.}\quad
\ii\Omega\leq\sigma.
\label{CPcondition}
\ee
One critical thing to note here is that in the fermionic case the two-sided bound
in \eqref{CPcondition} above implies that the
space of allowed states is   compact, whereas in the bosonic case the state space is unbounded.
  
In either case, the unitary Gaussian transformations can be seen as linear transformations on the system's quadrature/Majorana operators. And in either case these turn out to be the transformations that  preserve the system's (anti-)commutation relations. In the fermionic case these are orthogonal transformations (i.e. transformations that preserve the identity) and in the bosonic case they are symplectic transformations (i.e. transformations that preserve the symplectic form). An important difference between these groups is that the special orthogonal transformations form a compact group whereas the symplectic transformations do not.

Ultimately,  fermionic Gaussian dynamics is notably more restricted than bosonic dynamics. The fermionic state space is smaller in several ways: its first moment's all vanish (meaning no displaced states are possible), its covariance matrix is antisymmetric (which necessarily has less degrees of freedom than a symmetric matrix of the same dimension) and the state space itself is bounded/compact. As for the dynamics, comparing the fermionic partition performed here to the bosonic one performed in \cite{Dan2018} we find two less types of dynamics are possible. Furthermore, due to the compactness of the state space, fermionic Gaussian dynamics must either by cyclic or evolve to a fixed point, there is no infinite direction for the state to head off towards. This is in contrast to the bosonic case where the state may be squeezed, displaced or heated to an arbitrary degree without converging to a fixed point. 

\section{Conclusion}

We have introduced a classification of the generators of open fermionic Gaussian dynamics. Specifically we divided the generators the dynamics along four lines:
\begin{enumerate}
\item  unitary and non-unitary
\item  active and passive
\item  single-mode and multi-mode
\item  state-dependent and state-independent
\end{enumerate}
Of the potential sixteen types of dynamics expected of such a division, we find that seven of them vanish, leaving only nine types of fermionic Gaussian dynamics. 

We have provided illustrative examples of each of these types of dynamics. Our analysis of the complete positivity of these dynamics indicates that the presence of any non-unitary effects necessitates the presence of noise in the dynamics. Since this noise is active (it involved particle flux with the environment), completely positive fermionic Gaussian dynamics is either unitary or involve particle exchange with its environment.

We have also provided comparison with a similar partitioning of bosonic Gaussian dynamics \cite{Dan2018}. Overall, fermionic Gaussian states and transformation are more restricted than bosonic ones. For a finite number of modes, there are less degrees of freedom for both Gaussian states and transformations if the modes are fermionic as compared to if they are bosonic. As we discussed these restrictions ultimately stem from the system's (anti-)communtation relations.

Work that applies this partition to the dynamics of quantum systems that are bombarded by a rapid succession of fermionic ancillae is in progress.

\acknowledgments

E.M-M acknowledges support of the NSERC Discovery program and his Ontario Early Researcher award. This work was supported in part by the Natural Sciences and Engineering Research Council of Canada (NSERC).
D.G. acknowledges support by NSERC through the Vanier Scholarship. 

\appendix

\section{Converting Dynamics from Hilbert Space to Phase Space}\label{HilbertPhaseCrossover}
In this appendix we convert the system's evolution from the Heisenberg picture,
\begin{align}\label{AppHeisenbergEq}
\frac{\d}{\d t}\hat{\mathbf{r}} 
=\ii [\hat{H},\hat{\mathbf{r}}].
\end{align}
into a linear differential equation on the phase space vector, $\bm{r}$, in the case where the Hamiltonian is quadratic in the Majorana operators 
\begin{align}\label{Appeffect}
\hat{H} = \frac{\ii}{2} \bm{\hat{r}}^{\intercal}  \mathbf{H} \bm{\hat{r}}
\end{align}
for some $2N$ by $2N$ real-valued antisymmetric matrix $\mathbf{H}$.

Using \eqref{Appeffect} and distributing we find
\begin{align}
\frac{\d}{\d t}\hat{r}_{k} 
&= \ii [\hat{H},\hat{r}_{k}] 
\nonumber \\
&= -\frac{1}{2}[\sum^{2N}_{n,m=1} \mathbf{H}_{nm}\hat{r}_{n}\hat{r}_{m},\hat{r}_{k}]
\nonumber \\
&= -\frac{1}{2} \sum^{2N}_{n,m=1} \mathbf{H}_{nm} [\hat{r}_{n}\hat{r}_{m},\hat{r}_{k}]
\end{align}
and from here we can use the identity
\be
[\hat{A}\hat{B},\hat{C}]
=\hat{A}\lbrace \hat{B},\hat{C}\rbrace - \lbrace \hat{A},\hat{C}\rbrace \hat{B}
\ee
and \eqref{oform} to find
\begin{align}
\frac{\d}{\d t}\hat{r}_{k} 
&=-\frac{1}{2}  \sum^{2N}_{n,m=1} \mathbf{H}_{nm} \Big( \hat{r}_{n}\lbrace \hat{r}_{m},\hat{r}_{k}\rbrace - \lbrace \hat{r}_{n},\hat{r}_{k}\rbrace \hat{r}_{m}\Big)\\
\nonumber
&=-\frac{1}{2}\sum^{2N}_{n,m=1} \mathbf{H}_{nm} \Big(\hat{r}_{n}\delta_{mk} - \hat{r}_{m}\delta_{nk}\Big) \\
\nonumber
&=-\frac{1}{2} \sum_{n=1}^{2N} (\mathbf{H}_{nk} - \mathbf{H}_{kn})\hat{r}_{n}
\end{align}
where in the last step we have relabeled the indices in the second term. Since $\mathbf{H}$ is antisymmetric we have
\begin{align}
\frac{\d}{\d t}\hat{r}_{k} 
=\sum_{n=1}^{2N}\mathbf{H}_{kn}\hat{r}_{n}=(\mathbf{H}\hat{r})_k
\end{align}
or equivalently \cite{serge2011}
\begin{align}\label{Apprmaster}
\frac{\d}{\d t}\hat{\mathbf{r}} 
=\mathbf{H} \, \hat{\mathbf{r}}.
\end{align}
This is the result claimed in  \eqref{rmaster}.

\bibliography{references}
\end{document}